\documentclass[smallcondensed]{svjour3}    
\smartqed  
\usepackage{graphicx}
\usepackage{fix-cm}
\usepackage{bm}
\usepackage{amssymb}                                                                               
\usepackage{epsfig}
\usepackage{color}

\newcommand{\ket}[1]{\left| #1 \right\rangle}
\newcommand{\bra}[1]{\left\langle #1 \right |}

\newcommand{\spacer}{\rule[0cm]{0cm}{0cm}}

\newcommand{\Stab}{\mbox{\rm Stab}}

\newcommand{\diag}{\mbox{\rm diag}}

\newcommand{\R}{{\mathbb R}}

\newcommand{\C}{{\mathbb C}}

\newcommand{\Z}{{\mathbb Z}}

\newcommand{\Id}{{\rm Id}}

\newcommand{\twotwo}[4]{\left[ \begin{array}{cc} #1 & #2 \\
                        #3 & #4 \end{array} \right]}

\begin{document}

\title{{Local unitary group stabilizers and entanglement for multiqubit
    symmetric states}
\thanks{This work was supported by National Science Foundation grant
  \#PHY-0903690.}
}

\titlerunning{Local unitary group stabilizers of symmetric states}        

\author
{Curt~D.~Cenci\nolinebreak \and 
\nolinebreak David~W.~Lyons \nolinebreak \and \nolinebreak Scott~N.~Walck}

\authorrunning{Cenci, Lyons, and Walck} 

\institute{C.D. Cenci \at
              Mathematical Sciences, Lebanon Valley College, 101
              N. College Ave. Annville, PA 17003 \\
              \email{cdc002@lvc.edu}           
           \and
D.W. Lyons \at
              Mathematical Sciences, Lebanon Valley College, 101
              N. College Ave. Annville, PA 17003 \\
              \email{lyons@lvc.edu}           
           \and
S.N. Walck \at
              Physics, Lebanon Valley College, 101
              N. College Ave. Annville, PA 17003 \\
              \email{walck@lvc.edu}           
}

\date{Received: date / Accepted: date}
\maketitle

\begin{abstract}

We refine recent local unitary entanglement classification for symmetric
pure states of $n$ qubits (that is, states invariant under permutations
of qubits) using local unitary stabilizer subgroups and Majorana
configurations.  Stabilizer subgroups carry more entanglement
distinguishing power than do the stabilizer subalgebras used in our
previous work.  We extend to mixed states recent results about local
operations on pure symmetric states by showing that if two symmetric
density operators are equivalent by a local unitary operation, then they
are equivalent via a local unitary operation that is the {\em same} in
each qubit.  A geometric consequence, used in our entanglement
classification, is that two symmetric pure states are local unitary
equivalent if and only if their Majorana configurations can be
interchanged by a rotation of the Bloch sphere.

  \keywords{symmetric state \and
  local unitary \and stabilizer subgroup \and Majorana} \PACS{03.67.Mn}
\end{abstract}

\section{Introduction}
\label{intro}

The question of when a given multiparty state can be converted to
another by local operations and measurements of subsystems is crucial in
quantum information science~\cite{nielsenchuang}.  The fact that entangled
states play a role as resources in computation and communication
protocols motivates problems of measurement and
classification of entanglement.  In general, these are difficult
problems, already rich for the case of pure states of $n$-qubits, where
the number of real parameters necessary for classifying entanglement
types grows exponentially in $n$.

A promising special case for the general problem of entanglement
measurement and classification is that of the symmetric states, that is,
states of composite systems that are invariant under permutation of the
subsystems.  Symmetric states admit simplified analyses, and they
are of interest in their own right.  Examples of recent work in which
permutation invariance has made possible results where the general case
remains intractable include: geometric measure of entanglement
\cite{aulbach2010,aulbach2010b,markham2010}, efficient tomography \cite{toth2010},
classification of states equivalent under stochastic local operations
and classical communication (SLOCC) \cite{bastin2009,bastin2010}, and our
own work on classification of states equivalent under local unitary (LU)
transformations \cite{symmstatespaper}. 

The main result of this paper (Theorem~3 below) is a classification of
LU equivalence classes of $n$-qubit symmetric states that
refines our own previous work~\cite{symmstatespaper}, which is based on the
following idea.  Suppose states $\rho,\rho'$ are local unitary equivalent via
some LU transformation $U$, that is, we have $\rho'=U\rho U^\dagger$.
If a local unitary operator $V$ stabilizes $\rho$, then $UVU^\dagger$
stabilizes $\rho'$.  The consequence is that stabilizer subgroups of
locally equivalent states are isomorphic via conjugation.  Thus
the isomorphism class of the stabilizer is an LU invariant.  This
inspires a two-stage classification program.
\begin{enumerate}
\item Classify LU stabilizer subgroup conjugacy classes.
\item Classify LU classes of states for each of the stabilizer classes
  from stage~1.
\end{enumerate}
In previous
work~\cite{symmstatespaper,minorb1,minorb2,maxstabnonprod1,maxstabnonprod2,su2blockstates},
we have exploited the Lie algebra structure of the tangent space of
infinitesimal LU transformations, which is a linearization of the
stabilizer subgroup, to achieve results in both of these stages for
various classes of states.  A strength of this method is that linear Lie
algebra computations are more tractable than the corresponding nonlinear
Lie group computations.  The drawback is that a Lie algebra detects only
the connected component at the identity element of the corresponding Lie
group.  A stabilizer subalgebra does not ``see'' the discrete part of
the stabilizer subgroup.  For example, the Lie stabilizer subalgebra is
the zero vector space for most stabilizer states, that is, states
stabilized by the full $n$-qubit Pauli group. Group level information is
necessary to capture the local unitary stabilizer properties of such
states.

In Theorems~1 and~2 of~\cite{symmstatespaper}, we classify four infinite
families and 1 discrete family (that is, the zero vector space) of
stabilizer subalgebras for pure symmetric states, and identify LU
classes of pure symmetric states that have those stabilizers.  In this
paper, we advance this classification by separating the four infinite
families of subalgebras into six infinite families of groups,
inequivalent under isomorphism by local unitary conjugation.  The
discrete stabilizer subalgebra corresponds to finite subgroups of $SO(3)$
(rotations of real 3-dimensional Euclidean space): these are the
infinite family of cyclic groups; the infinite family of dihedral
groups; and the five finite symmetry groups of the Platonic solids.

Extending our previous stabilizer subalgebra classification for
symmetric states to a stabilizer sub{\em group} classification makes use
of the Majorana representation for pure symmetric states: Given a
collection $\ket{\psi_1},\ldots,\ket{\psi_n}$ of 1-qubit states, we can
symmetrize to form the state
$$\ket{\psi}=\alpha\sum_{\pi}
\ket{\psi_{\pi(1)}}\ket{\psi_{\pi(2)}}\cdots \ket{\psi_{\pi(n)}}$$ where
$\pi$ ranges over all $n!$ permutations of the $n$-qubits, and $\alpha$
is a normalization factor.  It is a fact (see~\cite{bastin2009}) that
{\em any} symmetric pure state can be written as such a symmetrization,
and further, that the set of $n$ 1-qubit states whose symmetrization is
$\ket{\psi}$ is unique up to phase factors.  Thus the set of symmetric
pure states is in one-to-one correspondence with configurations of
multisets (one or more of the 1-qubit states may be repeated) of $n$ of
points on the Bloch sphere.  

Using the fact that a rotation of the Bloch sphere corresponds to
unitary operation on 1-qubit states, it is a simple observation that a
rotation of the Majorana configuration of points representing a state
$\ket{\psi}$ results in an LU equivalent state $\ket{\psi'}=V^{\otimes
  n}\ket{\psi}$, where $V$ is the $2\times 2$ unitary operator
corresponding to the given rotation of the sphere.  Not obvious, but
true nonetheless, is that given {\em any} LU operation $U=U_1\otimes
U_2\otimes \cdots \otimes U_n$ that transforms a symmetric state
$\ket{\psi}$ to another symmetric state $\ket{\psi'}$, there is a
1-qubit operation $V$ such that $\ket{\psi'}=V^{\otimes n}\ket{\psi}$.
This was proved by Mathonet et al~\cite{mathonet2010} for SLOCC
operations on pure symmetric states.  We show in Theorem~1 below that
this holds more generally for LU operations on {\em mixed} symmetric states.
A consequence (Theorem~2) is that $\rho,\rho'$ are LU equivalent if and
only if their Majorana configurations can be interchanged by a rotation
of the Bloch sphere.  This fact is used in the proof of the main result
Theorem~3.

\section{Preliminaries}
\label{prelim}

We take $n$-qubit state space to be the set of $2^n\times 2^n$ density
matrices (positive semidefinite matrices with trace 1).  Pure states are
represented by density matrices of rank 1.  We write $\ket{D_n^{(k)}}$ to
  denote the Dicke state with $k$ excitations.

We take the local unitary group to be $PU(2)^n$,
where $PU(2)=U(2)/\{\lambda \; \Id\colon \lambda \in \C\,
|\lambda|=1\}$, called the
projective unitary group, is the set of projective equivalence classes
of matrices in $U(2)$.  That is, matrices $g,h$ in $U(2)$ represent the
same element in $PU(2)$ if and only if $g=\lambda h$ for complex number
$\lambda$.  The projective unitary group $PU(2)$ is 
isomorphic to the group $SO(3)$ of
rotations of 3-dimensional Euclidean space via
$$\lambda \exp\left(-i\theta/2 v\cdot \sigma\right) \leftrightarrow
\mbox{rotation by $\theta$ radians about axis $v$}
$$ where $\lambda$ is a norm 1 complex number, $\theta$ is a real
number, $v=(v_1,v_2,v_3)$ is a unit vector in $\R^3$, and
$\sigma=(\sigma_x,\sigma_y,\sigma_z)$ is the vector of Pauli matrices
(see \cite{nielsenchuang} Ex.~4.5.).

We will denote elements of $PU(2)$ and the local unitary group $PU(2)^n$
by their representatives $g$ in $U(2)$ and $U=(g_1,\ldots,g_n)$ in $U(2)^n$,
and will write $g\equiv h$ or $U\equiv V$ to denote equality in $PU(2)$
and $PU(2)^n$, and will use the equals sign to indicate equality in
$U(2)$ and $U(2)^n$.  Similarly, we will write
$\ket{\psi}\equiv\ket{\phi}$ to indicate that two state vectors are
equal up to phase.

The local unitary group element $U=(g_1,\ldots,g_n)$ acts on the density
matrix $\rho$ by
$$\rho \to U \rho U^\dagger = (g_1\otimes \cdots \otimes g_n)\; \rho \; (g_1\otimes \cdots
\otimes g_n)^\dagger.$$  We denote by $\Stab_\rho$ the local unitary stabilizer subgroup for
$\rho$ 
$$\Stab_\rho = \{U\in PU(2)^n\colon U\rho U^\dagger = \rho\}. $$

\section{Main Results}
\label{mainresults}

\begin{theorem}\label{luisblochrotation}
  Let $\rho, \rho'$ be $n$-qubit
  symmetric states, pure or mixed, with $n\geq 3$.  Then $\rho,\rho'$ are LU equivalent if
  and only if there exists an element $g$ in $U(2)$ such that 
$$\rho' =  (g^{\otimes n}) \rho (g^{\otimes n})^\dagger.$$
\end{theorem}

{\it Proof of Theorem~\ref{luisblochrotation}.}
We only need to prove the ``only if'' direction.  Let $\rho' = U\rho U^\dagger$, where
$$U=(g_1,g_2,\ldots,g_n)=\prod_{j=1}^ng_j^{(j)}$$
is an LU transformation.
Suppose there exist $k,\ell$ with $g_{k} \not \equiv g_{\ell}$.
Transposing the $k$-th and $l$-th coordinates of $U$, let
$$V= g_\ell^{(k)}g_k^{(\ell)} \prod_{j\neq k,\ell}g_j^{(j)}$$
By symmetry, we have $\rho' = V\rho V^\dagger$, and therefore we
have $\rho = (V^\dagger U) \rho (U^\dagger V).$

Let $h=g_{\ell}^{\dagger}g_k$ so that we have
$$V^\dagger U = h^{(k)} (h^\dagger)^{(\ell)},$$
and choose $u$ in $U(2)$ to diagonalize $h$, so that we
have 
$$uhu^{\dag} \equiv \left[ \begin{array}{cc} e^{it} & \\ & e^{-it}\end{array}
  \right]$$
for some $t$.  Let us call this diagonal matrix $d$, and 
let $\tau = (u^{\otimes n}) \rho (u^{\otimes n})^\dagger$, so that we have 
$$\tau = d^{(k)}(d^\dagger)^{(\ell)} \;\tau \; (d^\dagger)^{(k)}d^{(\ell)}.$$

Let $\tau = \sum_{IJ} c_{IJ} \ket{I}\bra{J}$ be the expansion of $\tau$ in
the computational basis, where $I=i_1\ldots i_n,J=j_1\ldots j_n$ denote binary strings of length
$n$ and the $c_{IJ}$ are complex coefficients.  We claim that if
$c_{IJ}\neq 0$, then $J=I$ or $J=I^c$, where $I^c$ denotes the bit
string obtained by taking the mod 2 complement of each bit in
the string $I$.  Suppose, on the other hand, that there exists a pair
$I,J$ such that $c_{IJ}\neq 0$ and $J\neq I$ and $J\neq I^c$.  Then
there exist two qubit labels $k,\ell$ such that $j_kj_\ell \neq
i_ki_\ell$ and $j_kj_\ell \neq
(i_ki_\ell)^c$.  Without loss of generality, suppose $(i_ki_\ell)=00$
and $(j_kj_\ell)=01$. Since
$$d^{(k)} (d^\dagger)^{(\ell)}
\ket{00}\bra{01}(d^\dagger)^{(k)}d^{(\ell)}=e^{-2it}\ket{00}\bra{01}$$ 
we must have $t=m\pi$ for some
integer $m$.  Then $d\equiv \Id$, and so $h\equiv \Id$, and therefore
$g_k \equiv g_\ell$, contradicting our assumption.
We conclude that 
\begin{equation}\label{ghzlikeform}
\tau = a\ket{I}\bra{I} + b\ket{I}\bra{I^c} + \overline{b}\ket{I^c}\bra{I}
+ (1-a)\ket{I^c}\bra{I^c}  
\end{equation}
for some coefficients $a,b$ and some bit string $I$.

Next we claim that we may assume $I=0\cdots 0$ or
$I=1\cdots 1$.  Suppose contrary 
that there are two qubit positions $k,\ell$ such that
$i_k\neq i_\ell$.  Choose any third qubit position $r$ (this is where we
use the hypothesis that $n\geq 3$).  We must have
$i_r=i_k$ or $i_r=i_\ell$.  Without loss of generality, suppose
$i_r=i_k$.  Now transpose qubits $\ell,r$.  This produces a state $\widetilde{\tau}$
with nonzero coefficient for the term $\ket{I'}\bra{I'}$, where
$i'_k=i'_\ell$.  But this contradicts the fact that
$\widetilde{\tau}=\tau$ because $\tau$ is symmetric.
This establishes the claim.

Next we claim that we may take $b$ to be real and nonnegative
in~(\ref{ghzlikeform}).  If $b$ is not real, let $\phi = \arg(b)/n$ if
$I=0\cdots 0$ and let $\phi=-\arg(b)/n$ if $I=1\cdots 1$.  Replacing
$\tau$ by $\left(\diag(1,e^{i\phi})\right)^{\otimes n}\; \tau \;
\left(\diag(1,e^{-i\phi})\right)^{\otimes n}$ establishes the claim.

Now apply the preceding argument to $\rho'$ to construct a sequence of
LU transformations that are the same in each qubit to obtain
$$\tau' = a'\ket{I'}\bra{I'} + b'\ket{I'}\bra{(I')^c} +
\overline{b'}\ket{(I')^c}\bra{I'} + (1-a')\ket{(I')^c}\bra{(I')^c}$$ for
some real and nonnegative coefficients $a',b'$ and some bit string
$I'=0\cdots 0$ or $I'=1\cdots 1$.  Comparing 1-qubit reduced density
matrices for $\tau,\tau'$ yields $a=a'$ or $a=1-a'$.  If the latter,
replace $\tau'$ by $(X,\ldots,X)\;\tau'\; (X,\ldots,X)$.  Finally,
comparing eigenvalues of $\tau,\tau'$, we conclude that $b=b'$.  Thus we
have constructed a chain of symmetric local unitary operations that
transform $\rho$ to $\rho'$, as desired.  This concludes the proof of
Theorem~\ref{luisblochrotation}.
 {\nopagebreak\spacer\hfill $\square$}



\begin{theorem}\label{luequivblochrot}
    Let $\ket{\psi}, \ket{\psi'}$ be $n$-qubit
  symmetric states with Majorana configurations ${\cal C}_\psi,{\cal
    C}_{\psi'}$.  Then $\ket{\psi},\ket{\psi'}$ are local unitary
  equivalent if and only if
there exists an element $g$ in $U(2)$ such that 
$${\cal C}_{\psi'}=g{\cal C}_\psi.$$
\end{theorem}

{\it Proof of Theorem~\ref{luequivblochrot}.}  Let
$\ket{\psi},\ket{\psi'}$ be symmetric states with Majorana
configurations ${\cal C}_\psi = \{\ket{\psi_1},\ldots,\ket{\psi_n}\}$ and
${\cal C}_{\psi'} = \{\ket{\psi'_1},\ldots,\ket{\psi'_n}\}$.  If there is a
rotation of the Bloch sphere given by $g$ in $U(2)$ that takes ${\cal C}_\psi$
to ${\cal C}_{\psi'}$, then (possibly after renumbering) we have
$g\ket{\psi_j}\equiv \ket{\psi'_j}$ for $1\leq j\leq n$, and hence
$\ket{\psi'}\equiv g^{\otimes n}\ket{\psi}$.  Conversely, if
$\ket{\psi'}=U\ket{\psi}$ for some local unitary $U$, then by Theorem~1,
there is a $g$ in $U(2)$ such that $\ket{\psi'}=g^{\otimes
  n}\ket{\psi}$.  We can interpret this $g$ as a rotation of the Bloch
sphere, and it is clear that we have ${\cal C}_{\psi'}=g{\cal C}_\psi$.
 {\nopagebreak\spacer\hfill $\square$}

\begin{theorem}\label{stabinf}
Let $\rho$ be an $n$-qubit symmetric pure state whose local unitary
stabilizer $\Stab_\rho$ is infinite.  Then one of the following holds.
\begin{enumerate}
\item [(i)] The state $\rho$ is LU equivalent to the product state
  $\tau=\ket{\psi}\bra{\psi}$, where $\ket{\psi}=\ket{0\cdots 0}$ and
  $\Stab_\rho$ is isomorphic to $U(1)^n$, where $(e^{it_1}, \ldots,
  e^{it_n})$ in $U(1)^n$ corresponds to
$$\left(\exp(-it_1Z/2),\dots,\exp(-it_nZ/2)\right)
$$
in $\Stab_\tau$.  There is one LU equivalence class of this type.
\item [(iia)] The state $\rho$ is LU equivalent to the GHZ state
  $\tau=\ket{\psi}\bra{\psi}$, where $\ket{\psi}=(1/\sqrt{2})(\ket{0\cdots 0} +
  \ket{1\cdots 1})$ for some $n\geq 3$ and $\Stab_\rho$ is isomorphic to $U(1)^{n-1}\rtimes
  \Z_2$, where $(e^{it_1}, \ldots, e^{it_{n-1}},b)$ in
  $U(1)^{n-1}\rtimes \Z_2$ corresponds to
$$\left(\exp(-it_1Z/2),\dots,\exp(-it_{n-1}Z/2),\exp(i\left(\sum_k
  t_k\right)Z/2)\right)\cdot (X,\ldots,X)^b
$$
in $\Stab_\tau$.  There is one LU equivalence class of this type.
\item [(iib)] The state $\rho$ is LU equivalent to the generalized GHZ
  state $\tau=\ket{\psi}\bra{\psi}$, where $\ket{\psi}=a\ket{0\cdots 0}
    + b\ket{1\cdots 1}$ for some $n\geq 3$ with $|a|\neq |b|$, and $\Stab_\rho$ is
    isomorphic to $U(1)^{n-1}$, where $(e^{it_1}, \ldots, e^{it_{n-1}})$
    in $U(1)^{n-1}$ corresponds to
$$\left(\exp(-it_1Z/2),\dots,\exp(-it_{n-1}Z/2),\exp(-i\left(\sum_k
  t_k\right)Z/2)\right)
$$
in $\Stab_\tau$.  We may take $a$ and $b$ to both be positive and real
with $a>b$.  The LU equivalence classes of this type are parameterized by
the interval $0<t<1$ by $a=\cos \frac{\pi}{4}t$, $b=\sin \frac{\pi}{4}t$.
\item[(iii)] The state  $\psi$ is LU equivalent to the singlet state
$\tau=\ket{\psi}\bra{\psi}$, where $\ket{\psi}=\ket{01} - \ket{10}$ 
and $\Stab_\rho$ is isomorphic to $PU(2)$, where 
$g$ in $PU(2)$ corresponds to
$$(g,g)
$$
in $\Stab_\tau$.  There is one LU equivalence class of this type.
\item [(iva)] The state $\psi$ is LU equivalent to the Dicke state
$\tau=\ket{\psi}\bra{\psi}$, where $\ket{\psi}=\ket{D_n^{n/2}}$ for some
  even $n\geq 4$,
and $\Stab_\rho$ is isomorphic to $U(1)\rtimes \Z_2$, where 
$(e^{it},b)$ in $U(1)\rtimes \Z_2$ corresponds to
$$\left(\exp(-itZ/2),\dots,\exp(-itZ/2)\right)\cdot (X,\ldots,X)^b
$$
in $\Stab_\tau$.  There is one LU equivalent class of this type.
\item [(ivb)] The state $\psi$ is LU equivalent to the Dicke state
  $\tau=\ket{\psi}\bra{\psi}$, where $\ket{\psi}=\ket{D_n^{k}}$ for some
  $n\geq 3$ and some $k$ in the range $0<k<n$ and $k\neq n/2$, and $\Stab_\rho$ is isomorphic to
  $U(1)$, where $(e^{it})$ in $U(1)$ corresponds to
$$\left(\exp(-itZ/2),\dots,\exp(-itZ/2)\right)
$$
in $\Stab_\tau$.  There are $\lfloor n/2\rfloor$ LU equivalence classes
of this type, with representatives
 $$\ket{D_n^{(1)}},\ket{D_n^{(2)}},\ldots,\ket{D_n^{(\lfloor{n/2}\rfloor
    - 1)}}.$$
\end{enumerate}
\end{theorem}

\paragraph{Proof of Theorem~\ref{stabinf}.}
We show in~\cite{symmstatespaper} that an arbitrary pure symmetric state
$\rho$ is LU equivalent to one of the states $\tau$ listed
in~(i)--(ivb).  Theorem~1 of that paper identifies 4 families of nonzero
stabilizer Lie subalgebras, which exponentiate to stabilizer subgroup
elements of the forms given in~(i)--(ivb).  In each case, it is easy to
see that the given correspondences are one-to-one.  To establish the
claimed isomorphisms, it remains to be shown 
that the groups
\begin{enumerate}
\item [(i)] $U(1)^n$
\item [(iia)] $U(1)^{n-1} \rtimes Z_2$
\item [(iib)] $U(1)^{n-1}$
\item [(iii)] $PU(2)$
\item [(iva)] $U(1) \rtimes \Z_2$
\item [(ivb)] $U(1)$
\end{enumerate}
given in (i)--(ivb) above map surjectively {\em onto} the full stabilizer subgroups
of the corresponding states given in (i)--(ivb).

Below we give the proof for~(iia) and~(iva).  The other proofs are both
similar and easier.  The proofs that the homomorphisms in~(iia)
and~(iva) are onto share the following outline.  We consider an
arbitrary element $U=(g_1,\ldots,g_n)$ in $\Stab_\tau$, and we wish to
show that $U$ is in the image of the given homomorphism. First, we show
that it suffices to show that that either all $g_k$ are diagonal, or all
$g_k$ are antidiagonal.  Then we consider two cases. The first case is
where $g_k\equiv g_\ell$ for all $k,\ell$, so that $U$ has the form
$U\equiv (g_1,\ldots,g_1)$ for some $g_1$ in $U(2)$.  The second case
with where there exists a pair of qubits $k,\ell$ such that
$g_k\not\equiv g_\ell$.  We show that both cases lead to the conclusion
that either all the $g_k$ are diagonal, or all the $g_k$ are
antidiagonal.  By the earlier reduction, this completes the proof.

\paragraph{Proof of surjectivity in Theorem~\ref{stabinf} $(iia)$.}
Let $\rho=\ket{\psi}\bra{\psi}$ be the $n$-qubit GHZ state, where
$\ket{\psi}=\ket{0\cdots 0}+\ket{1\cdots 1}$ for some $n\geq 3$, and let
$U=(g_1,\ldots,g_n)$ be an element of $\Stab_\rho$.  We aim to prove
that $U$ can be written in the form
$$\left(\exp(-it_1Z/2),\dots,\exp(-it_{n-1}Z/2),\exp(i\left(\sum_k
  t_k\right)Z/2)\right)\cdot (X,\ldots,X)^b
$$
for some real $t_1,\ldots,t_{n-1}$ and some $b=0,1$.

We begin with the claim that it suffices to show that either $g_k$ is
diagonal for all $k$, or $g_k$ is antidiagonal for all $k$.  Indeed, if
every $g_k$ is diagonal, say $g_k\equiv e^{it_kZ}$, then from
$$
U\ket{0\cdots 0}\bra{1\cdots 1}U^\dagger=\exp(2i(\sum t_k))\ket{0\cdots
  0}\bra{1\cdots 1}  
$$
we conclude that $\sum t_k$ is an integer multiple of $\pi$, and so
$\sum t_k$ may be taken to be zero (because we are working
projectively, we have $e^{i\pi Z}=-\Id\equiv \Id$).  Hence $U$ is the image of the element
$(e^{-it_1/2},\ldots,e^{-it_{n-1}/2})$ in $U(1)^{n-1}$.  If every $g_k$
is antidiagonal, then we can write
\begin{equation}\label{gkantidiag}
  g_k \equiv \twotwo{0}{e^{it_k}}{e^{-it_k}}{0} = \twotwo{e^{it_k}}{0}{0}{e^{-it_k}}X.
\end{equation}
Then from
$$
U\ket{0\cdots 0}\bra{1\cdots 1} U^\dagger =\exp(-2i\sum
t_k)\ket{1\cdots 1}\bra{0\cdots 0}
$$
we have $\sum t_k$ is $\pi$ times and integer, $U$ is projectively
equivalent to the image of 
$(e^{-it_1/2},\ldots,e^{-it_{n-1}/2},1)$ in $U(1)^{n-1}\rtimes \Z_2$.
This establishes the claim.

Next we show that either all $g_k$ are diagonal, or all $g_k$ are
antidiagonal.

{\bf Case a.} Suppose that $g_k\equiv g_\ell$ for all $k,\ell$.  Then
$U$ has the form $U\equiv (h,\ldots,h)$ for some $h$ in $U(2)$.  As in
the proof of Theorem~\ref{luequivblochrot}, we can
read $h$, and therefore $U$, as a rigid motion of the Bloch sphere, and
conclude that $U$ takes the Majorana configuration for the GHZ state
into itself.  Thus  $U$ is a symmetry
of the regular $n$-gon in the equatorial plane, and is therefore either
a rotation about the $Z$-axis, or a 180-degree-rotation about the
$X$-axis followed by a rotation about the $Z$-axis.  Thus $h$ is either
diagonal or antidiagonal.

{\bf Case b.} Suppose there exist qubits $k,\ell$ such that $g_k\not \equiv
g_\ell$.  As in the proof of Theorem~\ref{luisblochrotation}, let
$V=g_\ell^{(k)}g_k^{(\ell)}\prod_{j\neq k,\ell}g_j^{(j)}$, let
$h=g_\ell^\dagger g_k$, so that
$$V^\dagger U = h^{(k)}(h^\dagger)^{(\ell)}$$ is in $\Stab_\rho$.
Choose $u$ in $U(2)$ to diagonalize $h$, and let $\tau=u^{\otimes n}\;
\rho \; (u^{\otimes n})^\dagger$, so that we have
$d^{(k)}(d^\dagger)^{(\ell)}$ in $\Stab_\tau$, where $d = e^{it Z}\equiv
uhu^\dagger$, for some real $t$.  Continuing to follow the proof of
Theorem~\ref{luisblochrotation}, considering the action of
$d^{(k)}(d^\dagger)^{(\ell)}$ on qubits $k,\ell$, the presence of a
standard nonzero coefficient $c_{IJ}$ in the expansion of $\tau$ in the
computational basis with $J\neq I$ and $J\neq I^c$ in qubits $k,\ell$
leads to the contradiction that $g_k\equiv g_\ell$, so we conclude that
$\tau=\ket{\psi'}\bra{\psi'}$ where $\ket{\psi'}=a\ket{0\cdots 0}+
b\ket{1\cdots 1}$ is an LU-equivalent GHZ state with $|a|=|b|$.  The
Majorana configuration for $\tau$ is a regular $n$-gon in the equatorial
plane, so we may conclude that $u$ is a rigid motion of the Bloch sphere
that must be of the form $e^{i\phi Z}$ or $e^{i\phi Z} X$, so $u$ is
diagonal or antidiagonal.  From this we have that $h$ is diagonal, so
$g_k=g_\ell d_{\ell k}$ for some diagonal matrix $d_{\ell k}$.  It
follows that $U$ is of the form
$$U\equiv (g_1,
g_1 d_{12},\ldots,g_1 d_{1n}) = (g_1,\ldots,g_1)(1,\Id,
d_{12},\ldots,d_{1n}).$$  Since the action of
$(\Id,d_{12},\ldots,d_{1n})$ is a rotation about the $Z$-axis, and $U$
takes the Majorana configuration of $\tau$ to itself, it must be
that $g_1$ is a rigid motion of the Bloch sphere coming from either a
diagonal or antidiagonal matrix as in case~a, so we conclude that all
$g_k$ are diagonal or all $g_k$ are antidiagonal, as desired.

\paragraph{Proof of surjectivity in Theorem~\ref{stabinf} $(iva)$.}
Let $\rho$ be the Dicke state $\rho = \ket{\psi}\bra{\psi}$, where
$\ket{\psi}=\ket{D_n^{(n/2)}}$ for some even $n\geq 4$, and let
$U=(e^{it},g_1,\ldots,g_n)$ be an element of $\Stab_\rho$.  We aim to
prove that $U$ can be written in the form
$$\left(\exp(-itZ/2),\dots,\exp(-itZ/2)\right)\cdot (X,\ldots,X)^b
$$
for real $t$ and some $b=0,1$.

We begin with the claim that it suffices to show that either $g_k$ is
diagonal for all $k$, or $g_k$ is antidiagonal for all $k$.  Suppose
that all $g_k$ are diagonal, say $g_k \equiv e^{it_kZ}$.  Choose two
qubit labels $k,\ell$, choose a weight
$n/2$ multiindex $I=i_1i_2\ldots i_n$ such that $i_k=0,i_\ell=1$, and let
$J=j_1j_2\ldots j_n$ denote the multiindex that is formed by complementing the
$k$-th and $\ell$-th bits of $I$.  Then from 
$$
U\ket{I}\bra{J}U^\dagger =\exp(2i(t_k-t_\ell)\ket{I}\bra{J} 
$$
we conclude that $t_k-t_\ell$ is an integer multiple of $\pi$.  This
holds for all $k,\ell$, so we have $g_k\equiv g_1$ for all $k$, so that
$U\equiv(g_1,\ldots,g_1)$.
Thus
$U$ is the image of $(e^{-t_1/2},0)$ in $U(1)\rtimes \Z_2$.  If every
$g_k$ is antidiagonal, then again we may write $g_k$ in the form of
equation~(\ref{gkantidiag}).  Considering the action of $U$ on $\ket{I}\bra{J}$
above, the same argument goes through with minor changes, and we have
that $U$ is the image of $(e^{-t_1/2},1)$ in $U(1)\rtimes \Z_2$.
This establishes the claim.

Next we show that either all $g_k$ are diagonal, or all $g_k$ are
antidiagonal.

{\bf Case a.} Suppose that $g_k\equiv g_\ell$ for all $k,\ell$.  Then $U$
has the form $U=(h,\ldots,h)$ for some $h$ in $SU(2)$.  We can
read $h$ as a rotation of the Bloch sphere that must take the Majorana
configuration for $\ket{\psi}$ into itself.  Thus $h$ is a rotation about the
$Z$-axis, or $h$ is a 180-degree-rotation about the $X$-axis followed by a
$Z$-axis rotation.  In the first case, $h$ is diagonal.  In the second
case, $h$ is antidiagonal.

{\bf Case b.} Suppose there exist qubits $k,\ell$ such that $g_k\not
\equiv g_\ell$.  By the same argument as for case~b in the previous
proof of surjectivity for~(iia), we conclude that $\ket{\psi}$ is LU
equivalent to a state of the form $a\ket{0\cdots 0} + b\ket{1\cdots 1}$,
which is either a product state or a generalized GHZ state.  But this
violates the known LU classification (Theorem~1
of~\cite{symmstatespaper}) for symmetric states.  We conclude that
case~b cannot hold, and this ends the proof.
 {\nopagebreak\spacer\hfill $\square$}

\medskip

\section{Conclusion}
\label{conclusion}

We have completely classified LU equivalence classes of LU stabilizer
subgroups for pure symmetric states.  For infinite stabilizer subgroups, we have
given a complete classification of LU equivalence classes of symmetric
states.  For each finite stabilizer subgroup, there are an infinite number
of LU equivalence classes of symmetric states.  Each family is
characterized by a Majorana configuration, and the LU equivalent states
are precisely those whose Majorana configurations are obtained by
rotating the Bloch sphere.

In future work we hope to extend these results to mixed symmetric
states.  We are encouraged by the success of recent work
\cite{bastin2010} by Bastin et al., in which they extend to mixed
symmetric states their own SLOCC classification \cite{bastin2009} for pure symmetric
states.






\bibliographystyle{spphys}       

\end{document}